\documentstyle[psfig,pre,multicol,aps]{revtex}
\begin{document}

\setlength{\textwidth}{180mm} 
\setlength{\textheight}{240mm}
\setlength{\parskip}{2mm}

\title{Effective Equations for Photonic-Crystal Waveguides 
and Circuits}
\author{Sergei F. Mingaleev and  Yuri S. Kivshar}

\address{ Nonlinear Physics Group, Research School of Physical 
Sciences and Engineering \\ 
Australian National University, Canberra ACT 0200, Australia}

\maketitle

\begin{abstract}
We suggest a novel conceptual approach for describing the
properties of waveguides and circuits in photonic crystals,
based on the effective discrete equations that include the
long-range interaction effects. We demonstrate, on the example 
of sharp waveguide bends, that our approach is very effective and 
accurate for the study of bound states and transmission 
spectra of the photonic-crystal circuits, and disclose the 
importance of evanescent modes in their properties. 
\end{abstract}


\begin{multicols}{2}

\narrowtext


One of the most promising applications of photonic crystals is a
possibility to create compact integrated optical devices 
\cite{book_sakoda}, which would be analogous to integrated circuits 
in electronics, but operating entirely with light. 

Usually, the properties of photonic crystals and photonic-crystal
waveguides are studied by solving Maxwell's equations
numerically, and such calculations are time consuming. Moreover,
the numerical approach does not always provide a good physical
insight. The purpose of this Letter is to suggest a novel
approach, based on the effective discrete equations,  for
describing many of the properties of the photonic-crystal
waveguides and circuits, including the example of the transmission 
spectra of sharp waveguide bends. The effective discrete equations we 
derive below are somewhat analogous to the Kirchhoff equations for electric
circuits. However, in contrast to electronics, in photonic
crystals both diffraction and interference become important, 
and thus the resulting equations involve the long-range interaction 
effects.

To introduce our approach, we consider a two-dimensional (2D)
photonic crystal consisting of infinitely long dielectric rods
arranged in the form of a square lattice with the lattice spacing
$a$. We study the light propagation in the plane normal to the
rods, assuming that the rods have a radius $r_0=0.18a$ and the
dielectric constant $\varepsilon_0=11.56$ (this corresponds to
GaAs or Si at the wavelength $\sim 1.55$ $\mu m$). 
For the electric field $E(\vec{x}, t) = e^{- i \omega t} \,
E(\vec{x} \,|\, \omega)$ polarized parallel to the rods, 
Maxwell's equations reduce to the eigenvalue problem
\begin{equation}
\left[ \nabla^2  + \left( \frac{\omega}{c} \right)^2
\varepsilon(\vec{x}) \right] E(\vec{x} \,|\, \omega) = 0 \; ,
\label{eq-E-omega}
\end{equation}
which can be solved by the plane-wave method
\cite{Johnson:2001-173:OE}.  A perfect photonic crystal of this 
type possesses a large (38\%) complete band gap (between
$\omega=0.303 \times 2\pi c/a$ and $\omega=0.444 \times 2\pi
c/a$), and it has been extensively employed during last few years
for the study of bound states in waveguides and bends
\cite{Mekis:1998-4809:PRB}, transmission of light through sharp
bends \cite{Mekis:1996-3787:PRL,Lin:1998-274:SCI}, 
branches \cite{fan} and channel drop filters \cite{Fan:1998-960:PRL}, 
nonlinear localized modes in straight 
waveguides \cite{Mingaleev:2000-5777:PRE} and perfect photonic crystals
\cite{Mingaleev:2001-5474:PRL}. Recently, this type of photonic
crystal with a $90^o$ bent waveguide was fabricated experimentally
in macro-porous silicon with $a=0.57$ $\mu m$ and a complete band
gap at $1.55$ $\mu m$ \cite{Zijlstra:1999-2734:JVB}.

To create a waveguide circuit, we introduce a system of defects and assume, 
for simplicity, that the defects are identical rods of the radius
$r_d$ (with $\varepsilon_{d}$) located at the points $\vec{x}_m$,
where $m$ is the index number of the defect rods.  In the photonic
crystal with defects the dielectric constant
$\varepsilon(\vec{x})$ can be presented as a sum of the periodic
and defect-induced terms, i.e.
$\varepsilon(\vec{x})=\varepsilon_{p}(\vec{x})+
\varepsilon_{d}(\vec{x})$ , and, therefore, Eq. (\ref{eq-E-omega})
can be written in an integral form
\begin{equation}
E(\vec{x} \,|\, \omega) = \left( \frac{\omega}{c} \right)^2
\int d^2\vec{y} \,\,\, G(\vec{x},
\vec{y} \,|\, \omega) \, \varepsilon_{d}(\vec{y}) \,
E(\vec{y} \,|\, \omega) \; ,
\label{eq-green-int}
\end{equation}
where $G(\vec{x}, \vec{y} \,|\, \omega)$ is the Green function
(see, e.g., \cite{Mingaleev:2000-5777:PRE}).

The integral equation (\ref{eq-green-int}) can be solved
numerically in the case of a small number of the defect rods. 
However, such calculations become
severely restricted by the current computer facilities as soon as
we increase the number of the defect rods in order to create
photonic-crystal waveguides, waveguide bends, and branches
\cite{Mekis:1996-3787:PRL,Lin:1998-274:SCI,fan,Fan:1998-960:PRL}. 
Therefore, our primary goal in this Letter is to
develop a new approximate physical model that would allow the
application of fast numerical techniques combined with a
reasonable accuracy and the further possibility to study nonlinear
photonic crystals and waveguides.

When the defects support {\em monopole modes}, 
a reasonably accurate model can be derived by assuming that the electric 
field inside a defect rod remains constant. In this case, we can
average the electric field in the integral equation
(\ref{eq-green-int}) over the cross-section of the rods
\cite{Mingaleev:2000-5777:PRE,McGurn:1996-7059:PRB}, and derive an
approximate matrix equation for the amplitudes of the electric
field $E_n(\omega) \equiv E(\vec{x}_n \,|\, \omega)$ at the defect
sites,
\begin{eqnarray}
\sum_m M_{n,m}(\omega) E_m = 0 \; , \nonumber \\
M_{n,m}(\omega) = \varepsilon_{d} \,
J_{n,m}(\omega) - \delta_{n,m} \; ,
\label{eq-E-disc}
\end{eqnarray}
where $\delta_{n,m}$ is the Dirac's delta function, and
\begin{equation}
J_{n,m}(\omega) = \left( \frac{\omega}{c} \right)^2
\int_{r_d} d^2 \vec{y} \,\,\,
G(\vec{x}_n, \vec{x}_m + \vec{y} \,|\, \omega )
\label{Jnm}
\end{equation}
is a coupling constant determined through the Green function of a
perfect 2D photonic crystal 
\cite{Mingaleev:2000-5777:PRE,Mingaleev:2001-5474:PRL}.

\begin{figure}
\vspace*{0mm} 
\centerline{\hbox{
\psfig{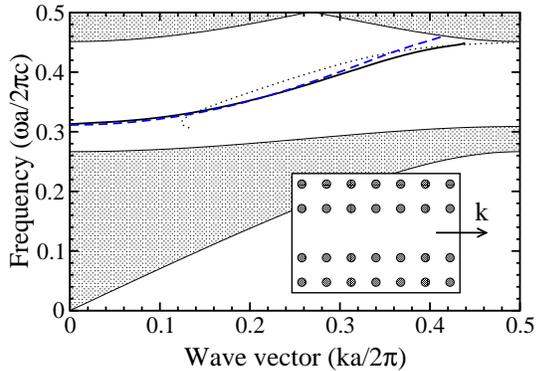}}}
\vspace*{2mm}
\caption{Dispersion relation for the 2D 
photonic-crystal waveguide (shown in the inset) calculated by the
super-cell method \protect\cite{Johnson:2001-173:OE} (dashed), and from
the approximate equations
(\ref{transfer-matrix})--(\ref{transfer-vect}) for $L$=7 (solid) and 
$L$=1 (dotted). The gray areas are the projected band structure 
of a perfect 2D crystal.}
\label{fig:dispersion}
\end{figure}

To check the accuracy of the approximate model (\ref{eq-E-disc}),
first we consider {\em a single defect} located at the point
$\vec{x}_0$. In this case,  Eq.~(\ref{eq-E-disc}) yields
$J_{0,0}(\omega_{d})= 1/\varepsilon_{d}$, and this expression
defines the frequency $\omega_{d}$ of the defect mode. For
example, applying this approach to the case when we have a defect 
created by a single removed rod, we
obtain the frequency $\omega_{d}=0.391 \times 2\pi c/a$ which
differs by only $1\%$ from the value $\omega_{d}=0.387 \times 2\pi
c/a$ calculated with the help of the MIT Photonic-Bands numerical
code \cite{Johnson:2001-173:OE}.

A single-mode waveguide can be created by removing a row of rods
(see the inset in Fig. \ref{fig:dispersion}). Assuming that the
waveguide is straight ($M_{n,m} \equiv M_{n-m}$) and neglecting
the coupling between asunder defect rods (i.e. $M_{n-m}=0$ for all
$|n-m|>L$), we rewrite Eq. (\ref{eq-E-disc}) in the
transfer-matrix form, $\vec{F}_{n+1}=\hat{T} \vec{F}_{n}$, where
we introduce the vector $\vec{F}_{n} = \{ \, E_n, \, E_{n-1}, \,
... \, , \, E_{n-2L+1} \, \}$ and the transfer matrix $\hat{T} =
\{ T_{i,j} \}$ with the non-zero elements
\begin{eqnarray}
&& T_{1,j}(\omega)= -\frac{M_{L-j}(\omega)}{M_{L}(\omega)}
\quad \mbox{for} \quad j=1, 2, ... , 2L \; , \nonumber \\
&& T_{j,j+1}=1 \quad \mbox{for} \quad j=1, 2, ... , 2L-1 \; .
\label{transfer-matrix}
\end{eqnarray}
Solving the eigenvalue problem
\begin{eqnarray}
\hat{T}(\omega) \vec{\Phi}^{p} = \exp\{i k_p(\omega)\} \,
\vec{\Phi}^{p} \; ,
\label{transfer-vect}
\end{eqnarray}
we can find the $2L$ eigenmodes of the photonic-crystal
waveguide. The eigenmodes with real wavenumbers $k_p(\omega)$
correspond to the propagating waveguide modes. In the waveguide
shown in Fig. \ref{fig:dispersion},  there exist only two such
modes (we denote them as $\vec{\Phi}^{1}$ and $\vec{\Phi}^{2}$),
propagating in the opposite directions ($k_1=-k_2>0$). In Fig.
\ref{fig:dispersion} we plot the dispersion relation $k_1(\omega)$
found from Eq. (\ref{transfer-vect}) for the nearest-neighbor
interaction ($L$=1) and  also taking into account interaction between 
seven neighbors 
($L$=7); we compare the results with those calculated directly by the
super-cell method \cite{Johnson:2001-173:OE}. As soon as we go
beyond the approximation of the nearest neighbors and take into
account the coupling between several defect rods, Eqs.
(\ref{eq-E-disc})--(\ref{transfer-vect}) provide {\em very
accurate results} for the dispersion characteristics of the
photonic-crystal waveguides. We verify that this conclusion is also 
valid for multi-mode waveguides, e.g. those created by removing 
several rows of rods.

\begin{figure}
\vspace*{0mm} 
\centerline{\hbox{
\psfig{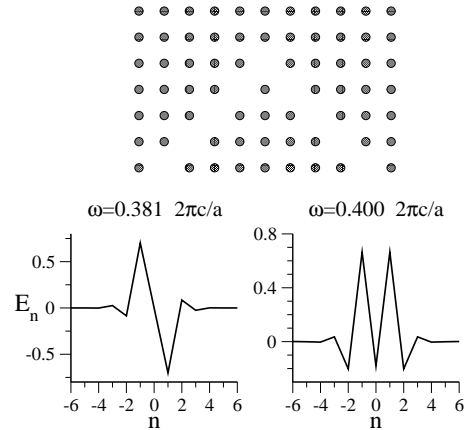}}}
\vspace*{2mm}
\caption{Electric field $E_n$ for two bound
states supported by a $90^o$ waveguide bend (shown in the top).
Center of the bend is located at $n=0$.}
\label{fig:bound}
\end{figure}

In addition to the propagating guided modes, in photonic-crystal
waveguides there always exist {\em evanescent modes} with
imaginary $k_p$. These modes, which cannot be accounted for in the
framework of the nearest-neighbor approximation,  remain somewhat
``hidden'' in straight waveguides, but they become important in
more elaborated structures such as waveguide bends and branches.
Importantly, our model does take into account all such effects.

We consider the simplest case of a waveguide bend, where the
evanescent modes manifest themselves in two different ways. First
of all, they create {\em localized bound states} in the vicinity
of the bend. As was shown in Ref. \cite{Mekis:1998-4809:PRB}, in
the cases when the waveguide bend can be considered as a finite
section of a waveguide of different type, the bound states
correspond closely to cavity modes excited in this finite section.
However, such a simplified one-dimensional model does not
describe correctly more complicated cases, even the bent waveguide
depicted in Fig. \ref{fig:bound} \cite{Mekis:1998-4809:PRB}. The 
situation becomes even more complicated for the waveguide branches
\cite{fan}.  In contrast, solving Eq. (\ref{eq-E-disc}) we can
find the frequencies and profiles of the bound states excited in
an arbitrary complex set of defects. As an example, in Fig.
\ref{fig:bound} we plot the profiles of two bound states (cf. Fig.
9 in Ref. \cite{Mekis:1998-4809:PRB}). The frequencies of the
modes are found from Eq. (\ref{eq-E-disc}) with the accuracy of
$1.5\%$.

Additionally, the evanescent modes determine the non-trivial
transmission properties of the waveguide bends which can also be
calculated with the help of our discrete equations. To demonstrate
this, we consider a bent waveguide consisting of two coupled
semi-infinite straight waveguides with a finite section 
(an arbitrary complex set of defects) between them.  
The finite section includes a
bend with a safety margin of the straight waveguide at both ends.
We assume that the defect rods inside this segment are
characterized by the index that runs from $a$ to $b$, and the
amplitudes $E_m$ ($m=a, ... , b$) of the electric field near the
sites of the removed rods are all unknown. We number the guided
modes (\ref{transfer-vect}) in the following way:  $p=1$
corresponds to the mode propagating in the direction of the
waveguide bend (for both ends of the waveguide), $p=2$ corresponds
to the mode, propagating in the opposite direction, $p=3, ... ,
L+1$ correspond to the evanescent modes which grow in the
direction of the bend, and $p=L+2, ... , 2L$ correspond to the
evanescent modes which decay in the direction of the bend. Then,
we can write the incoming and outcoming waves in the semi-infinite
waveguide sections as a superposition of the guided modes:
\begin{eqnarray}
E^{in}_m &=& \Phi^1_{a-m} + r \Phi^2_{a-m} +
\sum_{p=3}^{L+1} \lambda^{in}_p \Phi^p_{a-m} \; ,
\label{waves-in}
\end{eqnarray}
for $m=a-2L, ... , a-1$, and
\begin{eqnarray}
E^{out}_m &=& t \Phi^2_{m-b} +
\sum_{p=3}^{L+1} \lambda^{out}_p \Phi^p_{m-b} \; ,
\label{waves-out}
\end{eqnarray}
for $m=b+1, ... , b+2L$, where $\lambda^{in}_p$ and
$\lambda^{out}_p$ are unknown amplitudes of the evanescent modes
growing in the direction of the bend, and $t$ and $r$ are unknown
amplitudes of the transmitted and reflected propagating waves. We
take into account that the evanescent modes growing in the
direction from the bend vanish,  and assume that the amplitude of
the incoming plane wave $\vec{\Phi}^1$ is normalized to the unity.
Now, substituting Eqs. (\ref{waves-in})--(\ref{waves-out}) into
Eq. (\ref{eq-E-disc}), we obtain a system of linear equations with
$2L+b-a+1$ unknown. Solving this system, we find the transmission
$|t|^2$ and reflection $|r|^2$ coefficients.

\begin{figure}
\vspace*{0mm}
\centerline{\hbox{
\psfig{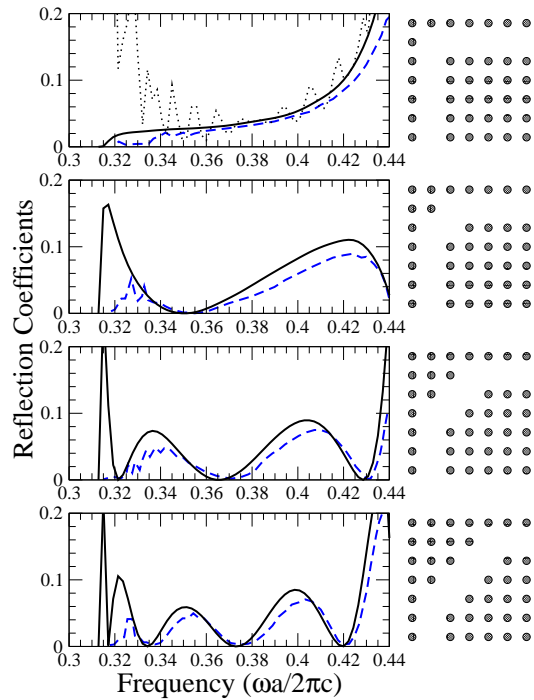}}}
\vspace*{2mm}
\caption{Reflection coefficients calculated by
the finite-difference time-domain method (dashed, from Ref. \protect\cite{Mekis:1996-3787:PRL}) and
from Eqs. (\ref{eq-E-disc})--(\ref{transfer-vect}) with $L=7$
(full lines) and $L=1$ (dotted, only in the top plot), for four
different bend geometries.}
\label{fig:scatt}
\end{figure}

In Fig. \ref{fig:scatt} we present our results for the
transmission spectra of several types of bent waveguides, as in
Ref. \cite{Mekis:1996-3787:PRL},  where the possibility of high
transmission through sharp bends in photonic-crystal waveguides
was first demonstrated. As is clearly seen, Eqs.
(\ref{eq-E-disc})--(\ref{waves-out}) provide a very accurate
method for calculating the transmission spectra of the waveguide
bends.
 
In conclusion, we have suggested a novel conceptual approach for
describing the properties of photonic-crystal waveguides and 
circuits, including the transmission spectra of sharp bends. The
effective discrete equations we have introduced here emphasize the
important role of the evanescent modes in the photonic-crystal
circuits, and they can be applied to study more complicated
problems such as transmission in waveguide branches, channel drop
filters, nonlinear localized modes in nonlinear waveguides, and so on.

The authors are indebted to S.H. Fan for useful comments 
and to A. Mekis for providing the data from Ref.~\cite{Mekis:1996-3787:PRL}. 
The work has been partially supported by the Australian Research Council.

\vspace*{-5mm}

\end{multicols}
\end{document}